\documentclass[pra,showpacs,aps,twocolumn,superscriptaddress]{revtex4-1}

\usepackage{amsmath, amssymb}
\usepackage{graphicx}
\graphicspath{{./plots/}}

\newcommand{\tr}{\mathop{\mathrm{tr}}\nolimits}

\begin{document}
\title{Edge-state blockade of transport in quantum dot arrays}

\author{M\'onica Benito}
\affiliation{Instituto de Ciencia de Materiales de Madrid, CSIC, 28049 Madrid, Spain}
\author{Michael Niklas}
\affiliation{Institut f\"ur Theoretische Physik, Universit\"at Regensburg, 93040 Regensburg, Germany}
\author{Gloria Platero}
\affiliation{Instituto de Ciencia de Materiales de Madrid, CSIC, 28049 Madrid, Spain}
\author{Sigmund Kohler}
\affiliation{Instituto de Ciencia de Materiales de Madrid, CSIC, 28049 Madrid, Spain}

\date{\today}


\begin{abstract}
We propose a transport blockade mechanism in quantum dot arrays and
conducting molecules based on an interplay of Coulomb repulsion and the
formation of edge states.  As a model we employ a dimer chain that exhibits
a topological phase transition.  The connection to a strongly biased electron
source and drain enables transport.  We show that the related emergence of
edge states is manifest in the shot noise properties as it is accompanied
by a crossover from bunched electron transport to a Poissonian process.  For both regions we
develop a scenario that can be captured by a rate
equation.  The resulting analytical expressions for the Fano factor agree
well with the numerical solution of a full quantum master equation.
\end{abstract}

\maketitle

\section{Introduction}

Quantum electronics is governed by charging energies
which give rise to Coulomb blockade which is apparent in
the diamond-like charging diagrams of quantum dots \cite{vanderWiel2003a} 
and conducting molecules
\cite{CuevasScheer2010}.  When electron spins and phonons
come into play, additional blockade phenomena may influence
the current-voltage characteristics.  For example, the Pauli exclusion
principle may cause a spin blockade in
double \cite{Weinmann1995a, Ono2002a} and triple quantum dots
\cite{Busl2013a}.  Moreover, in suspended quantum dots, an entering electron
may emit a phonon and become trapped until it reabsorbs a phonon, which is
known as a phonon blockade \cite{Weig2004a}.

Some blockade phenomena are less pronounced in the current, but have a
strong impact on the current noise.  Most prominently, the strong coupling
of an electron in a molecular wire with a vibrational degree of freedom may
lead to a switching between conducting and almost isolating configurations
and cause Franck-Condon blockade.  Then the transport becomes avalanchelike,
which drastically enhances the shot noise \cite{Koch2005a, Leturcq2009a}.
A similar effect occurs in capacitively coupled transport channels, where
noise measurements reveal that a mutual channel blockade causes electron
bunching \cite{Barthold2006a, Sanchez2008a}.

A one-dimensional tight-binding model with alternating tunnel matrix
elements represents a simple description of a dimerized polymer
\cite{Su1979a}.  It is characterized by a topological invariant, the Zak
phase \cite{Zak1989a}, which depends on the ratio between the inter- and
intradimer coupling and has been measured recently \cite{Atala2013a}.  For
finite chains in the topologically nontrivial phase, a pair of
exponentially decaying edge states emerges
\cite{Delplace2011a}.  Moreover, Coulomb interaction may lead to long-range
tunneling of doublons between edge states \cite{Bello2015a}.
When the chain is in contact with the electron source and drain, however, the
impact of the edge states on the transport properties remains an open
question.

\begin{figure}
\centerline{\includegraphics{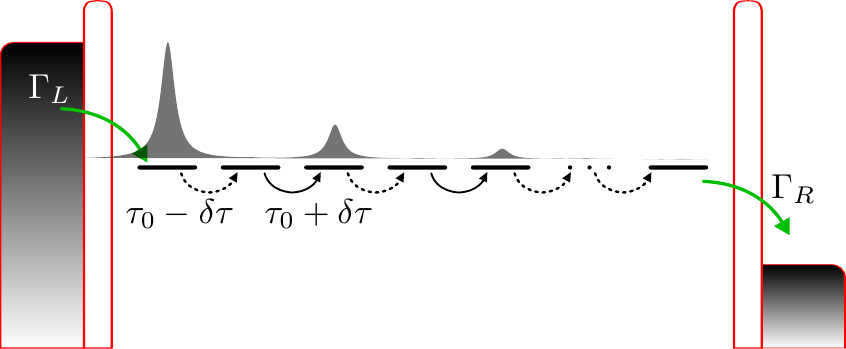}}
\caption{Dimer chain with tunnel couplings
$\tau'=\tau_0-\delta\tau$ and $\tau=\tau_0+\delta\tau$, respectively,
connected to the electron source (left) and drain (right).  At $\delta\tau=0$,
the chain undergoes a topological phase transition. The wave function
depicts the stationary state in the topological regime. Electron trapping in
the edge state at the source causes an edge-state blockade.}
\label{fig:setup}
\end{figure}

In this paper we propose an \textit{edge-state current blockade} in voltage-biased arrays such as that sketched in Fig.~\ref{fig:setup}, which relates to
the transition from a topologically trivial to a nontrivial regime.  We
show that it is most clearly visible in the shot noise.  In
Sec.~\ref{sec:model}, we introduce our model and a master equation
description.  The main results and the physical mechanism of the resulting
transport are presented in Sec.~\ref{sec:results}.  Finally, in
Sec.~\ref{sec:experiment} we discuss possible experimental realizations and draw our
conclusions in Sec.~\ref{sec:conclusions}.  Some technical aspects of the
numerical scheme and details of the calculations can be found in the Appendixes.

\section{Model and master equation}
\label{sec:model}

We employ spinless electrons on an array of length $N$ described by the
Hamiltonian $H_0 = H_\text{SSH} + H_\text{int}$.  It contains nearest-neighbor tunneling according to the Su-Shrieffer-Heeger (SSH) Hamiltonian
\cite{Su1979a}
\begin{equation}
H_\text{SSH} = \sum_{n=1}^{N-1} \tau_n c_{n+1}^\dagger c_{n} + \text{H.c.},
\end{equation}
with the alternating tunnel matrix elements $\tau_n = \tau_0
+(-1)^n\delta\tau$ and the fermionic annihilation operator $c_n$.  We keep
$\tau_0$ constant and use $\delta\tau$ as a control parameter.

\begin{figure}
\centerline{\includegraphics{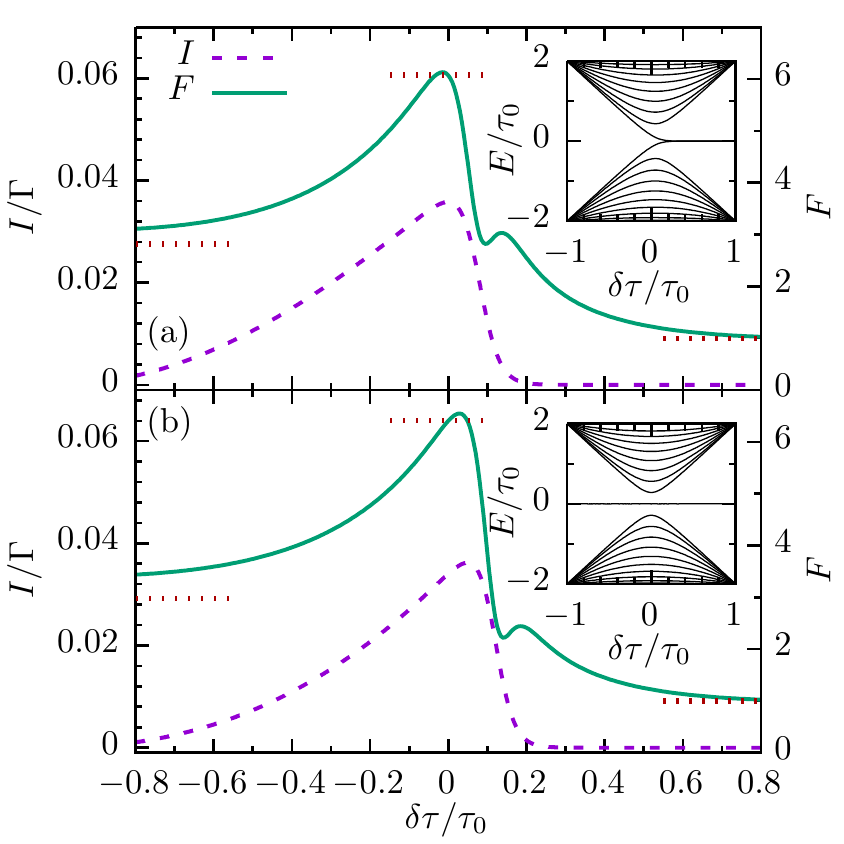}}
\caption{Current (dashed line) and Fano factor $F=C_2/|I|$ (solid line) for an
array of (a) $N=20$ and (b) $N=21$ sites as a function of the imbalance
$\delta\tau/\tau_0$ and the lead couplings $\Gamma_R=\Gamma_L=5\tau_0$.
The dotted horizontal lines mark the analytically obtained limits.  Despite
the different single-particle spectra (insets), the results
for an even and odd number of sites are qualitatively the same.}
\label{fig:IF}
\end{figure}

The SSH model is probably the simplest one with a topological phase
transition.  For $\delta\tau<0$, it describes a chain of weakly coupled
dimers which form two bands with a gap that closes at $\delta\tau=0$. When
$\delta\tau$ assumes positive values, two edge states emerge [see the inset of
Fig.~\ref{fig:IF}(a)].  In the bulk, the wave function of the edge states
decays exponentially with a localization length given by the inverse of
$\kappa = \ln(\tau/\tau') \approx 2\delta\tau/\tau_0$.  Thus, for
finite arrays, the edge states form a doublet with a level splitting
$\Delta \approx \tau_0 \exp({-N\delta\tau/\tau_0})$ (see Appendix
\ref{app:overlap}).  It will turn out that this doublet governs the
transport properties for $\delta\tau>0$.  If the
array consists of an odd number of sites, a monomer  will remain 
forming an edge state.  Thus, we witness
a transition from a situation with an edge state at the right end
of the chain ($\delta\tau<0$)  to one with an edge state at the left end
($\delta\tau>0$) \cite{Bernevig2013}. This transition, however, is not visible in the
spectrum [see inset of Fig.~\ref{fig:IF}(b)].

For the Coulomb repulsion, we assume $H_\text{int} =
\sum_{n>n'}U_{|n-n'|}N_n N_{n'}$ with the site occupations $N_n$ and the
interaction energies $U_{d}$ which decay with the distance $d=n-n'$ between
the sites.  Moreover, by working with spinless electrons, we have already ruled
out double occupation of a single site.  Physically, this is justified by
the typically very strong on-site interaction $U_0$ in quantum dots.

To enable transport, we couple the ends of the array to biased leads acting
as the electron source and drain with a voltage bias $V$.  Within second-order
order perturbation theory we integrate out the leads to obtain a Bloch-Redfield type master equation for the
reduced density operator. For
low temperatures and in the limit $\tau,\tau' \ll eV \ll U_d < U_0$, only
single-electron states are energetically accessible and the electron
transport becomes unidirectional.  Moreover, the array-lead tunneling
becomes independent of the details of the array's level structure. Then the
master equation assumes the convenient Lindblad form
\begin{equation}
\label{ME}
\dot\rho = \mathcal{L}\rho \equiv -\frac{i}{\hbar}[H_\text{SSH},\rho]
+ \Gamma_L \mathcal{D}(c_1^\dagger)\rho +\Gamma_R \mathcal{D}(c_N)\rho ,
\end{equation}
with $\mathcal{D}(x)\rho = (2x \rho x^\dagger - x^\dagger x\rho -
\rho x^\dagger x)/2$ and the dot-lead rates $\Gamma_{L,R}$.
The first term in $\mathcal{D}(x)$ corresponds to incoherent transitions
induced by the operator $x = c_1^\dagger,c_N$, which in our case is the
electron tunneling from the source to the array and from the array to the
drain, respectively.  Thus, the (particle) current is described by the
superoperator $\mathcal{J}\rho = \Gamma_L c_1^\dagger \rho c_1$ (or
alternatively by $\Gamma_R c_N \rho c_N^\dagger$).
Notice that neither the bias $V$ nor the interaction constant $U$ appear
explicitly in Eq.~\eqref{ME}. Let us therefore emphasize that our master
equation holds only in the limit in which strong Coulomb repulsion inhibits
the occupation with two or more electrons, i.e., it has to be evaluated in
the subspace of zero or one electrons on the chain.  As a consequence,
the dynamics on the chain is governed by the single-particle quantum
mechanics induced by the SSH Hamiltonian, while the electron tunneling
from the source to the chain is affected by the interaction.

Low-frequency current fluctuations can be characterized by the counting
statistics of the transported electrons.  For this purpose, we introduce a
counting variable $\chi$ and consider the modified master equation $\dot
R_\chi = \mathcal{L}_\chi R_\chi$ with $\mathcal{L}_\chi = \mathcal{L} +
(e^{i\chi}-1)\mathcal{J}$ \cite{Bagrets2003a, Novotny2004a}.  It is
constructed such that $\tr(R_\chi) = \langle e^{i\chi N_R}\rangle$ becomes
the moment generating function for the electron number in the drain,
$\phi(\chi,t)$.  The current cumulants $C_n = (\partial/\partial{i\chi})^n
\ln\dot{\phi}(\chi,t)|_{\chi=0,t\rightarrow\infty}$ contain the full
information about the low-frequency noise.  The spectral decomposition of
$R_\chi$ into the eigenbasis of $\mathcal{L}_\chi$ yields a formal
solution which at long times is dominated by the eigenvalue
with the largest real part, $\lambda_0(\chi)$.  Then,
$R_\chi \propto \exp[\lambda_0(\chi)t]$ and, thus, $\ln\phi(\chi,t) =
\lambda_0(\chi)t$.  Being interested in derivatives close to
$\chi=0$, we can treat $\chi$ as a small parameter and obtain 
the cumulants from an iteration based
on the Rayleigh-Schr\"odinger perturbation theory \cite{Flindt2008a}.  The
first two steps yield the current $I = C_1 = \tr(\mathcal{J}\rho_0)$ and
the variance $C_2 = I - 2\tr(\mathcal{J}\mathcal{R}\mathcal{J}\rho_0)$
\cite{Novotny2004a}, where $\rho_0$ is the stationary solution of the
master equation \eqref{ME} and $\mathcal{R}$ is the pseudo inverse of
$\mathcal{L}$.  For details, see Appendix~\ref{app:FCS}.

It is worthwhile to define the Fano factor $F = C_2/|I|$, which is a
dimensionless measure of the noise strength and hints at the nature of the
transport mechanism \cite{Blanter2000a}.  The value $F=1$ corresponds to
uncorrelated events, while larger values indicate bunching.  For more
profound statements, one has to consider also cumulants of higher order.

\section{Edge states, current, and shot noise}
\label{sec:results}

\subsection{General scenario for dimer chains}

Let us start by investigating a dimer chain, i.e., the case of an even
number of sites for which the current in the different regimes is shown in
Fig.~\ref{fig:IF}(a).  We notice that in the monomer limit
$\delta\tau=0$, the current assumes an appreciable value.  Towards both
the topologically trivial and the nontrivial region, it decays.
In the nontrivial region, the decay is faster despite the presence of
interband states.  The asymmetry is also found for the
Fano factor which is super-Poissonian for $\delta\tau\lesssim 0$, while for
$\delta\tau>0$ it converges to the Poissonian value $F=1$.  This indicates
that the transport relates to topology.

To reveal the physics behind this observation, we conjecture for each
region a dominating mechanism and capture it by a rate equation
that provides analytical expressions for the current and the Fano factor.
For the monomer chain realized at the transition
point $\delta\tau=0$ (for finite systems it is rather a crossover at
$\delta\tau\approx \tau_0/N$ \cite{Delplace2011a}), the eigenstates read
$\phi_\ell(n) \propto \sin[\pi\ell n/(N+1)]$, where $\ell = 1,\ldots,N$,
labels the solutions.  We assume that
each eigenstate forms a transport channel, where a strong Coulomb interaction
leads to mutual exclusion of the channel occupation.  The corresponding
load and unload rates $\gamma_\ell^{L,R}$ are determined by the overlaps with the
terminating sites, i.e., by $|\phi_\ell(1)|^2$ and $|\phi_\ell(N)|^2$.  For
a symmetric setup, $\gamma_\ell^L=\gamma_\ell^R\equiv \gamma_\ell$.
States with $\ell\approx N/2$ are much stronger coupled to the leads than those
with $\ell=1$ or $\ell=N$ and, thus, most of the time, the strongly coupled
states support a regular current.  However, whenever a weakly coupled state
becomes populated, an electron will remain there for the rather long time
$\gamma_\ell^{-1}$ and thereby interrupt the transport process.
Accordingly, we expect bunching as is indicated by a large Fano factor.
For a quantitative treatment, we formulate the above scenario as a rate
equation from which we obtain the current $I= \Gamma/(N+1)$ and the Fano
factor $F_\text{mono}(N) \approx(N-2)/3$.  Since the effects
are most noticeable in longer arrays, we ignore corrections of the
order $N^{-1}$. For the full expressions and their derivation, see
Appendix~\ref{app:channels}.

Deep in the trivial region $\delta\tau<0$, the central system
consists of weakly coupled dimers.  Then we can consider each dimer as one
site and, thus, expect the behavior of a monomer array with $N/2$ sites.
Therefore, without an explicit calculation, we can conclude that the Fano factor
is $F = F_\text{mono}(N/2)$.

Finally, in the topological region $\delta\tau>0$, the electrons mainly enter
and leave the array via an edge state which is at zero energy.  Since all other
states are energetically far off, they merely mediate long-range tunneling with the
exponentially small effective matrix element $\Delta$ given above.
This means that the situation can be captured by a two-level system.  For a
sufficiently large array, $\Delta\ll\Gamma$, the bottleneck of the
transport is the tunneling between edge states.  The corresponding current
reads $I\simeq\Delta^2/\Gamma$ and consists of uncorrelated events
\cite{Kaiser2006a}, i.e., it is a Poissonian process with the characteristic
Fano factor $F=1$.  For an explicit derivation, see Appendix
\ref{app:twosite}.

The Fano factor of the full numerical calculation agrees rather well with
the limits obtained analytically [see the horizontal lines in
Fig.~\ref{fig:IF}(a)].  This provides evidence that the transport process in
each region indeed follows the scenario sketched above.

\begin{figure}
\includegraphics{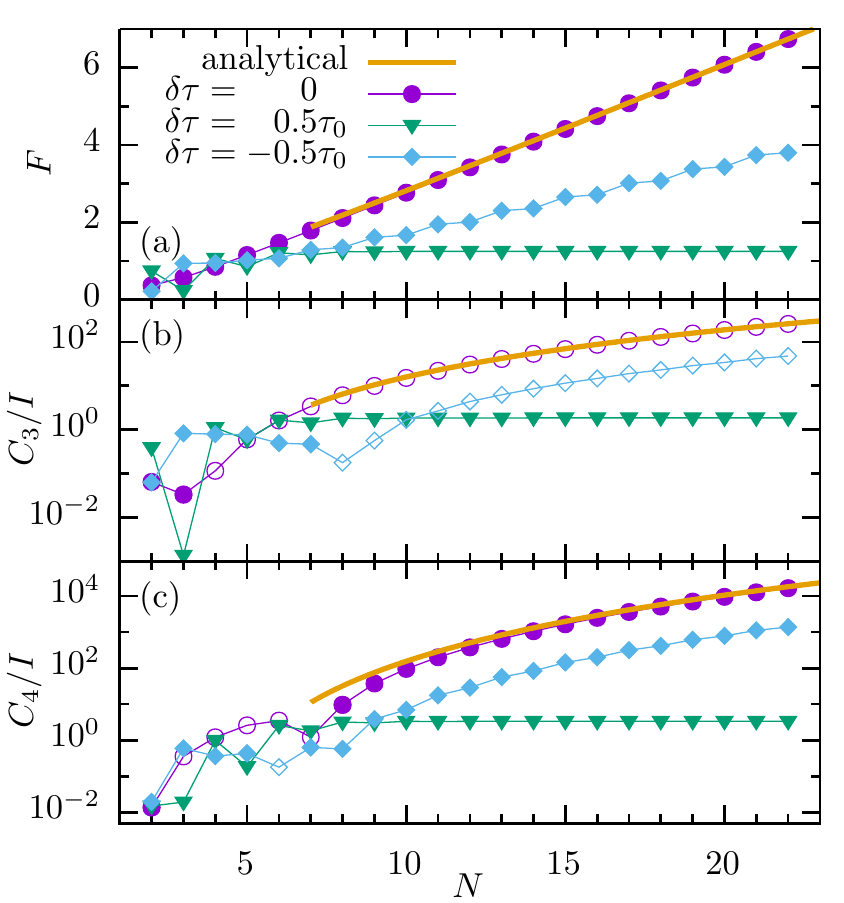}
\caption{(a) Fano factor, (b) third cumulant and (c) fourth cumulant as a
function of the chain length for various $\delta\tau$ and the lead coupling
$\Gamma_R=\Gamma_L=5\tau_0$.  Solid symbols mark positive values and stroked
symbols correspond to negative values.
}
\label{fig:F(N)}
\end{figure}
Since the separation of the Fano factors in the different regions grows
with the length of the array, one may aim at an experimental realization
with as many sites as possible.  This, however, will raise the experimental
difficulties drastically.  Moreover, beyond a certain system size, the
limit of a strong Coulomb blockade may no longer be realistic.  Thus the
length dependence of the Fano factors deserves a closer inspection.  The
data shown in Fig.~\ref{fig:F(N)}(a) confirm our analytical results even
down to rather small lengths.  For an intermediate length $N\approx 10$,
the Fano factors in the three regimes are already significantly different
from each other. In particular, the differences are larger than the
demonstrated resolution of mesoscopic noise measurements
\cite{Kiesslich2007a}.  The data for cumulants of higher order presented in
Figs.~\ref{fig:F(N)}(b) and \ref{fig:F(N)}(c) support our conjecture of
Poissonian transport in the topological phase.

\subsection{Arrays with an odd number of sites}

A further important observation is that the behavior of the shot noise for
chains with an odd number of sites interpolates the behavior of dimer chains.  In
particular, we find that the current and the Fano factor as a function of
$\delta\tau$ indeed are qualitatively the same as for even $N$ [see
Fig.~\ref{fig:IF}(b)].

For odd $N$, irrespective of the sign of $\delta\tau$, there always exists
one edge state which has zero energy [see the spectrum shown in the inset
of Fig.~\ref{fig:IF}(b)].  Thus, the chain does not exhibit a transition
between a topological and a nontopological phase.  Nevertheless, the
emergence of the edge state at one specific end of the chain can be
explained in terms of the bulk-edge correspondence as follows.  Let us
consider a not too short chain with even $N$ and $\delta\tau>0$, such that
the tunnel splitting $\Delta \sim \exp(-N\delta\tau/\tau_0)$ between the
edge states is much smaller than
the lead coupling $\Gamma$.  Then decoherence will turn a possible
superposition of both edge states into a mixture so that the edge state at
the source will not be influenced by its counterpart at the drain.  Then
removing the last site of the chain will not have a major effect on the
edge-state formation at the source.  In this sense, also finite chains with
odd $N$ still exhibit some footprint of a topological transition that is
found for infinite or semi-infinite dimer chains.

The common feature for even and for odd $N$ is that only for
$\delta\tau>0$, does the chain possess an edge state at the electron source.
The relevance of its location at the source is visible in the behavior
under inverting the applied bias: For even $N$, the chain is symmetric, so
that only the direction of the current changes.  Therefore, the Fano factor
in Fig.~\ref{fig:IF}(a) will remain the same.  For odd $N$, by contrast,
the inverted bias leads to a situation with an edge state at the drain but
none at the source.  Thus, bias inversion is equivalent to changing the
sign of $\delta\tau$, which for odd $N$ moves the edge state from one
end of the chain to the other.  Therefore, upon bias inversion, $F$ in
Fig.~\ref{fig:IF}(b) becomes reflected at the $y$ axis (not shown).

\subsection{Blocking mechanism and localization}

\begin{figure}
\includegraphics{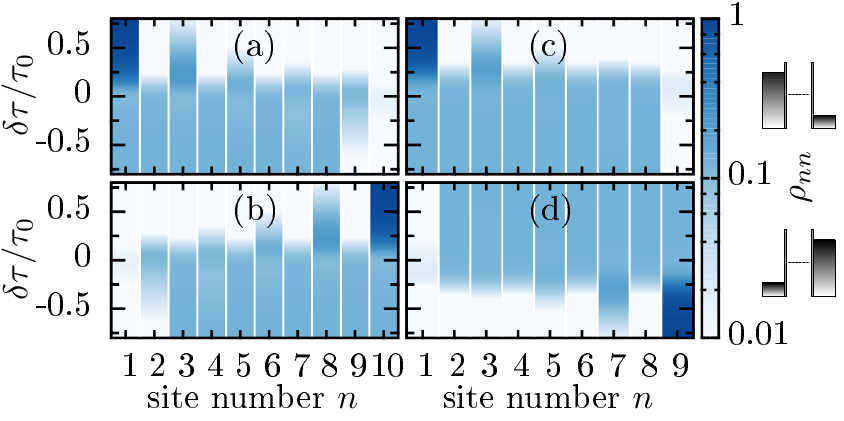}
\caption{Population of the quantum dots in the stationary state for the
array lengths (a),(b) $N=10$ and (c),(d) $N=9$ and the lead coupling
$\Gamma_R=\Gamma_L=5\tau_0$.  The data in the lower row are with the source and
drain interchanged, as indicated by the sketches at the right margin.
They reveal that a current blockade emerges when the edge state at the source
is strongly populated (dark blue areas).
Comparing the upper row with the lower row highlights the reflection
symmetry for even $N$, while for odd $N$ the spatial reflection corresponds to
inverting the sign of $\delta\tau$.
}
\label{fig:Pn}
\end{figure}

To underline the importance of the edge state and to develop a physical
picture for the blockade, we consider the population of the sites in the
stationary state of the open system [see Fig.~\ref{fig:Pn}].  For an even
number of sites [Figs.~\ref{fig:Pn} (a) and~\ref{fig:Pn}(b), where the latter is computed with
source and drain interchanged], in the topological phase ($\delta\tau>0$) the edge state at the source is predominantly populated.  This is
consistent with the scenario drawn above in which the transport occurs via
weak long-range tunneling.  Consequently, an electron becomes trapped in
the edge state localized at the source, while once it is at the opposite
side of the array, it leaves quickly to the drain.

For an odd number of sites, the behavior is similar. Outside the crossover
region $|\delta\tau|\gg\tau_0$, one edge state always exists.  For
$\delta\tau>0$, it is localized at site $1$ and causes a current blockade
[see Fig.~\ref{fig:IF}(b)].  By contrast, for $\delta\tau<0$, despite the
emergence of an edge state at site $N$, an appreciable current flows.

To resolve this seeming contradiction, let us focus on an array with odd
$N$ and $\delta\tau<0$ such that an edge state at the drain is formed.
Nevertheless, a small overlap of the bulk states with the last site opens a
way to circumvent the edge state.  Moreover, in rare cases in which
an electron reaches the edge state, it will proceed quickly to the drain,
consequently, no relevant blockade occurs.  For $\delta\tau>0$, the edge state
is located at the source and is mostly occupied [see Fig.~\ref{fig:Pn}(c)].
Then, bypassing site~1 is in principle possible, but would require double
occupation of the chain.  This, however, is inhibited by Coulomb repulsion so
that transport is interrupted until the electron in the edge state is
released.  This reveals that the blockade results from an interplay
of edge-state formation at the source and strong Coulomb repulsion.
The population for interchanged source and drain [Fig.~\ref{fig:Pn}(d)]
confirms that the edge-state formation at the source is also decisive
for trapping an electron when $N$ is odd.

\subsection{Disorder}
The formation of edge states with exponentially small splitting is
protected by sublattice symmetry present in our idealized array
Hamiltonian $H_\text{SSH}$.  In a realistic experiment, however, it may be quite
difficult to tune the system sufficiently well.  To investigate the
influence of imperfections, we consider disorder and add random on-site
energies,
\begin{equation}
H_\text{SSH}\to H_\text{SSH}+W\sum \xi_n c_n^\dagger c_n,
\end{equation}
where $W$ is the disorder strength and $\xi_n$ is taken from a normalized
box distribution with $-1/2\leq\xi_n\leq 1/2$.

\begin{figure}
\includegraphics{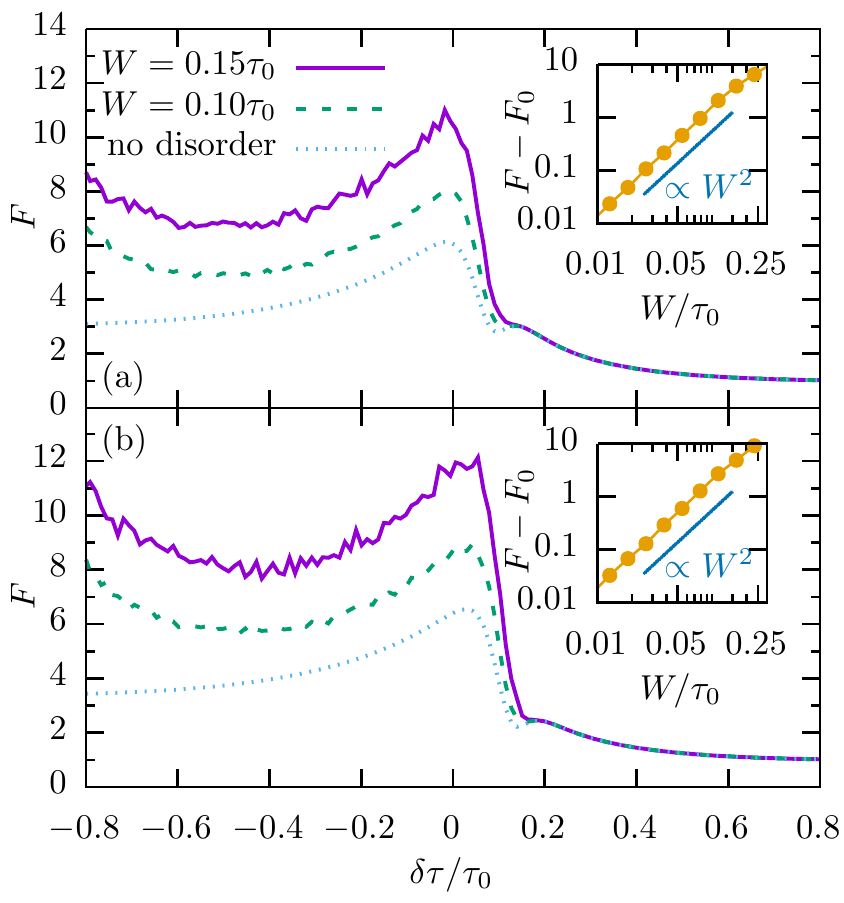}
\caption{Fano factor in the presence of disorder with strength $W$ for a
chain of lengths (a) $N=20$  and (b) $N=21$  with the parameters used in
Fig.~\ref{fig:IF}.
Insets: Deviation of the averaged Fano factor from its value in the absence
of disorder for $\delta\tau=-0.5\tau_0$.}
\label{fig:disorder}
\end{figure}

Figure~\ref{fig:disorder} shows the resulting Fano factor, now defined as
$\bar C_2/\bar I$, i.e., the ratio of the averages.  Comparing Figs.~\ref{fig:disorder}(a)
and~\ref{fig:disorder}(b), the behavior for an even and an odd number of
sites again turns out to be practically the same.
For $\delta\tau\lesssim 0$, we find that the Fano factor grows with
increasing disorder.  The enhancement is roughly $\propto W^2$, as can be
appreciated in the inset.  Notice that for larger values of $W$ and much
longer arrays, Anderson localization \cite{Anderson1958a} becomes relevant
and may change this behavior.

For $\delta\tau>0$, by contrast, disorder has almost no influence on
the Fano factor.  This finding is consistent with the physical picture
drawn above: The transport occurs via the two states localized at the ends
of the array, while the other states are off-resonant and not populated.
Since disorder even supports localization, the Poissonian behavior remains
unaffected.

\section{Possible experimental realization}
\label{sec:experiment}

The high tunability of the various types of quantum dots makes them
natural candidates for the implementation of blockade effects in mesoscopic
transport.  Recently, two parallel quantum dot arrays, each with seven dots, have
been demonstrated \cite{Puddy2015a}. In such systems, the charging and the
tunnel matrix elements are highly controllable by gate voltages. Thus it
should be possible to tune them such that they meet the requirement of an
interaction much larger than the tunneling, at least in not too long
arrays.

Molecular wires represent a realistic alternative, in particular, since they
are rather small and thus possess huge charging energies.  Between
experimental runs, they can be modified by atomic force microscopy
techniques \cite{Kocic2015a}.  Since this may also affect wire-lead
tunneling rates, the visibility of the blockade in the Fano factor is a
virtue since this quantity, in contrast to the current, depends only weakly
on the wire-lead coupling.  Moreover, one may change the topology of the
molecule by ac fields \cite{GomezLeon2013a}.

\section{Conclusions}
\label{sec:conclusions}

We have investigated a current blockade mechanism for strongly biased
contacted dimer chains.  It results from an interplay of Coulomb repulsion
and edge-state formation which relates to a topological transition.  The
edge state at the source can trap an electron, while Coulomb repulsion
inhibits a further electron to enter the chain.  The resulting electron
transport consists of rare tunnel events between the edge states and
exhibits a characteristic Poissonian behavior.  By contrast, in the
topologically trivial region, we find transport through delocalized states
and electron bunching.  Since the edge state at the source turned out to be
responsible, the effect can be observed also in chains with an odd number
of sites in which a different but related transition occurs, namely, the
displacement of the edge state from one end to the other.  Clear
experimental evidence for the transition between the different regions can
be provided by shot noise measurements.  While we have demonstrated that
the mechanisms on both sides of the transition are fairly insensitive to
static disorder, a more realistic description of an implementation with
molecular wires should consider also spin effects, vibrational degrees of
freedom, and decoherence.

\begin{acknowledgments}
We would like to thank \'Alvaro G\'omez-Le\'on for inspiring discussions.
This work was supported by the Spanish Ministry of Economy and
Competitiveness via Grant No.\ MAT2014-58241-P and by the DFG via SFB~689.
\end{acknowledgments}

\appendix
\section{Overlap of the edge states}
\label{app:overlap}

The Schr\"odinger equation for a dimer chain with intra- and interdimer
couplings $\tau$ and $\tau'$, respectively, is
\begin{equation}
 \tau\sigma_+\phi_{n-1}+ \tau'\sigma_x\phi_n +\tau\sigma_-\phi_{n+1}
 =\epsilon_n \phi_n,
\end{equation}
where $\phi_n=\left(c_{2n},c_{2n+1}\right)^T$ and $n$ labels the unit
cells.  For periodic boundary conditions we use the Bloch
ansatz $\phi_n = e^{ikn}\varphi(k)$ and obtain the Bloch equation
\begin{equation}
\label{app:bloch}
\begin{pmatrix}
0 & \tau e^{-i k}+\tau'  \\
\tau e^{ik}+\tau' & 0  
\end{pmatrix} \varphi(k) = \epsilon(k)\varphi(k).
\end{equation}
An edge state in a semi-infinite chain corresponds to a solution that
vanishes at some site such that, e.g., $\phi_{-1}=0$.  Then, we obtain from
the Schr\"odinger equation and Eq.~\eqref{app:bloch} the condition
\begin{equation}
\label{app:edge}
\begin{pmatrix}
0 & \tau'  \\
\tau e^{ik}+\tau' & 0  
\end{pmatrix} \varphi(k) = 0.
\end{equation}
It possesses a nontrivial solution if $k = \pi+i\ln(\tau/\tau')$, which for
$\tau>\tau'$ is decaying as $\phi_n\propto \exp(-\kappa n)$ with the
exponent $\kappa = \ln(\tau/\tau')$.  Close to the phase transition
$|\delta\tau|\ll\tau_0$, it becomes $\kappa = 2\delta\tau/\tau_0$.  Therefore,
the overlap between the two edge states of a chain with $N/2$ dimers can be
estimated as
\begin{equation}
\label{Delta}
\Delta \approx \tau_0 e^{-\delta\tau N/\tau_0}.
\end{equation}
It agrees with the splitting of the interband doublet found in finite
dimer chains \cite{Delplace2011a}.

\section{Iteration scheme for the cumulants}
\label{app:FCS}

As we are interested in the statistics of the transport, we need to
generalize the master equation formalism, introducing a counting variable
$\chi$ which keeps track of the electron number in the leads. The cumulants
of the corresponding distribution function are given by the $k$th
derivatives with respect to $i\chi$ at $\chi=0$ of the logarithm of the
moment generating function $\phi(\chi,t)=\langle e^{i\chi N_R}\rangle$.
The moment generating function can be written as the trace of
the generalized reduced density operator $R_\chi(t)=\tr_\text{leads}
(\rho_\text{tot} e^{i\chi N_R})$, which obeys the
master equation
\begin{equation}
\dot{R}_\chi(t) = \mathcal{L}_\chi R_\chi(t),
\end{equation}
where $\mathcal{L}_\chi=\mathcal{L}+\left(e^{i\chi}-1\right)\mathcal J$.
Notice that we have restricted ourselves to unidirectional transport, i.e.,
to the limit of large
bias in which all relevant eigenstates of the conductor are
within the voltage window and thermal excitations do not play
a role.

In the long-time limit, the dynamics of $R_\chi(t)$ is governed by the
eigenvalue of $\mathcal{L}_\chi$ with the largest real part, denoted as
$\lambda_0(\chi)$. Then, $R_\chi(t) \propto \exp[\lambda_0(\chi)t]$ and,
thus, $\ln\phi(\chi,t) = \lambda_0(\chi)t$ (besides a correction that
vanishes in the long-time limit).  Instead of calculating the proper
eigenvalue of $\mathcal{L}_\chi$ and its derivatives with respect to
$\chi$, one can treat $\chi$ as a small parameter and obtain the cumulants
from an iteration based on Rayleigh-Schr\"odinger perturbation theory
\cite{Flindt2008a,Dominguez2010a}.  The solution in the Markovian case is
\begin{equation}
\label{Ck}
C_k = \sum_{k'=0}^{k-1}\binom{k}{k'}\tr (\mathcal{J}P_{k'}),
\end{equation}
where $C_0=0$ and $\mathcal{L} P_0=0$.  The other components $P_k$ follow
from the equation
\begin{equation}
\label{Pk}
\mathcal{L} P_k = -\sum_{k'=0}^{k-1}\binom{k}{k'}
(\mathcal{J}-C_{k-k'})P_{k'} \,,
\end{equation}
which has to be solved under the condition $\tr P_k=0$.  This step is
equivalent to applying the pseudoinverse of the Liouvillian to the
right-hand side of Eq.~\eqref{Pk}.  In this way, the first cumulant, i.e.,
the current, can be written as $C_1=\tr(\mathcal{J} P_0)$. This enables the
computation of $P_1$ from the equation $\mathcal{L}
P_1=-(\mathcal{J}-C_{1})P_0$.  Then the second cumulant, i.e.,
the zero-frequency noise, becomes $C_2=C_1+2\tr(\mathcal{J} P_1)$.

\section{Analytical approach to the transport cumulants}
\label{app:analytics}

\begin{figure}
\includegraphics{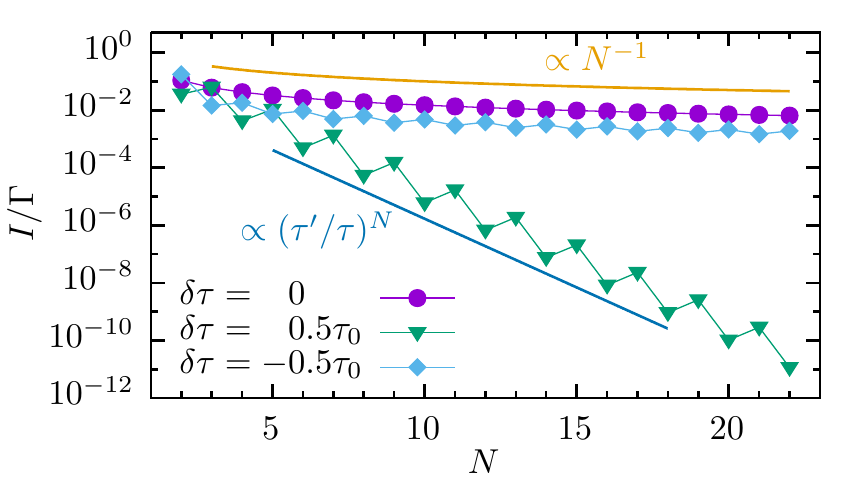}
\caption{Stationary current as a function of the chain length for the values
of $\delta\tau$ displayed.  The dot-lead coupling is $\Gamma_R=\Gamma_L=5\tau_0$.
}
\label{fig:I(N)}
\end{figure}

The current for the full model follows directly from the stationary
solution of the master equation \eqref{ME} of the main text, i.e., from the
kernel of the Liouvillian $\mathcal{L}$.  It can be computed analytically,
which allows us to evaluate the expression for the current.  For an even
number of sites, we obtain
\begin{equation}
\label{Inumeven}
I_\text{even} = \frac{\Gamma_R}{N+\frac{\Gamma_R}{\Gamma_L}+\frac{\Gamma_R^2}{4\tau^2}\left[N-2+\left(\frac{\tau}{\tau'}\right)^N\right]} ,
\end{equation}
while for odd $N$, the current reads
\begin{equation}
I_\text{odd} = \frac{\Gamma_R}{\frac{\Gamma_R}{\Gamma_L}+\frac{\Gamma_R^2\left(N-1\right)}{4\tau^2}+\left(\frac{\tau'}{\tau}\right)^{2}\left[N-1+\left(\frac{\tau}{\tau'}\right)^{N+1}\right]} .
\end{equation}
Both expressions assume their maximum close to $\tau\approx\tau'$.  For
$\tau\gg\tau'$, i.e., in the region in which we find edge-state blockade,
it decays $\propto (\tau'/\tau)^N$.  In the opposite limit, $\tau\ll\tau'$,
the decay is algebraic, $I\propto N^{-1}$ (see Fig. \ref{fig:I(N)}).

By contrast, computing the cumulants $C_n$ with $n\geq 2$ requires not only
the kernel of the Liouvillian, but also its pseudoinverse, which considerably
complicates the analytical solution.  To nevertheless find
analytical results for the noise, below we develop a description with a simplified master equation for the two limits
discussed in the main text.

\subsection{Mutually exclusive channels}
\label{app:channels}

\begin{figure}
\includegraphics{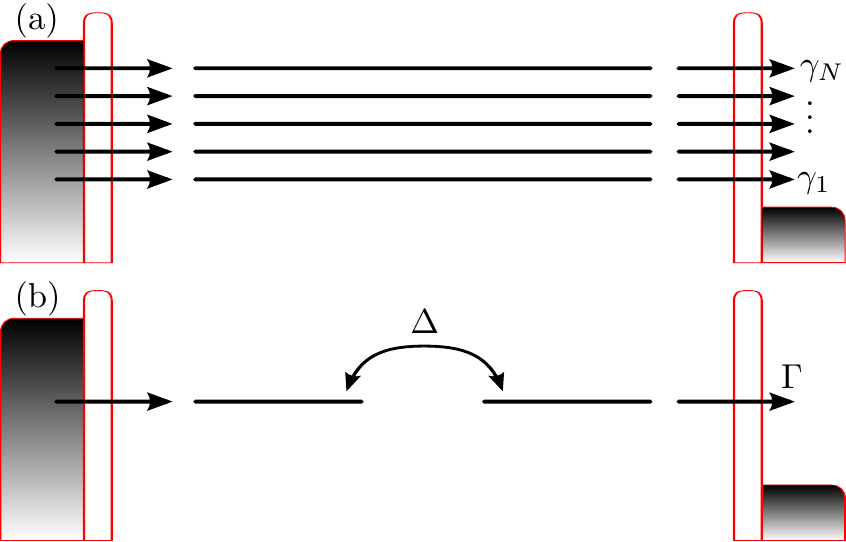}
\caption{Sketch of the situations that we treat analytically with
rate equations.
(a) Mutually exclusive channels for the delocalized eigenstates of a
monomer chain.  The rates $\gamma_\ell$ reflect the overlap between the
eigenstates and the first and the last site and obey $\sum_\ell\gamma_\ell
= \Gamma$.
(b) Two-state model for the edge states in the topological region.  The
intersite tunneling $\Delta$ is the exponentially small overlap between
the edge states given in Eq.~\eqref{Delta}.}
\label{fig:fig6}
\end{figure}

A general model for transport via mutually exclusive channels $\ell$ that are weakly
coupled to both leads with equal strength is sketched in Fig.~\ref{fig:fig6}(a).
It corresponds to the rate equation
\begin{equation}
\label{app:me.monomer}
\dot P = \begin{pmatrix}
-\Gamma & \gamma_1 & \ldots & \gamma_N \\
\gamma_1 & -\gamma_1 &  & 0 \\
\vdots &   & \ddots & \vdots \\
\gamma_N & 0 & \ldots & -\gamma_N
\end{pmatrix}
\begin{pmatrix}
p_0 \\ p_1 \\ \vdots \\ p_N \end{pmatrix} ,
\end{equation}
where normalization is ensured by $\Gamma = \sum_\ell\gamma_\ell$.
The rates $\gamma_\ell$ are determined by the overlap between the eigenstates
$\phi_\ell$ with the terminating sites.  In a symmetric setup, the rates at
the source and at the drain are equal, which is reflected by the symmetry
of the matrix in Eq.~\eqref{app:me.monomer}.
To be specific, for $\delta\tau=0$ the eigenstates of the array are
\begin{equation}
\phi_\ell=\sqrt{\frac{2}{N+1}}\sin\Big(\frac{\pi \ell n}{N+1}\Big) ,
\end{equation}
so that the rates become
\begin{equation}
\label{app:gamma}
\gamma_\ell = \frac{2\Gamma}{N+1}\sin^2\Big(\frac{\pi\ell}{N+1}\Big).
\end{equation}
Then the stationary solution of Eq.~\eqref{app:me.monomer} reads
$P_0=(1,1,...,1)^T/(N+1)$ and thus $I=\Gamma/(N+1)$, which represents the
weak coupling limit of Eq.~\eqref{Inumeven}.

The second cumulant follows from evaluating the formal solution derived
above.  It reads
\begin{equation}
C_2 = I +\frac{2\Gamma}{(N+1)^3}
      \bigg[\frac{\Gamma}{\tilde\Gamma} - N(N+1)\bigg] ,
\end{equation}
where $\tilde\Gamma^{-1} = \sum_\ell \gamma_\ell^{-1}$ is dominated by the
weakly coupled states owing to their small $\gamma_\ell$.  Inserting the
rates and performing the iteration scheme also for the next two orders, we
find
\begin{align}
\frac{C_2}{I} ={}&  \frac{N^2-N+3}{3\left(N+1\right)} 
\equiv F_{\text{mono}}(N) ,
\\
\frac{C_{3}}{I} ={}&   -\frac{N^2(N-7)}{30}+O(N) ,\\
\frac{C_{4}}{I} ={}&   \frac{N^4(2N-25)}{315}+O(N^3).
\end{align}
Notice that the cumulant ratio grows with the length of the
array as $C_{n+1}/C_n \propto N^2$.

\subsection{Two-site model}
\label{app:twosite}

In the topological region and for a sufficiently long array, the transport
occurs mainly via long-range tunneling from one edge state to the other,
while the population of the other eigenstates is negligible.  Then a proper
simplified model is that of a two-level system with tunnel splitting
$\Delta$ and a coupling to the source and drain, as is sketched in
Fig.~\ref{fig:fig6}(b).  It can be captured by the
master equation (in the basis $\{|0\rangle\langle 0|,|L\rangle \langle
L|,|R\rangle \langle R|,|L\rangle \langle R|,|R\rangle \langle L|\}$)
\begin{equation}
\dot \rho = \begin{pmatrix}
-\Gamma_L & 0 & \Gamma_R & 0 & 0 \\
\Gamma_L & 0 & 0 & i \Delta/2 &  -i\Delta/2 \\
 0 & 0 &  -\Gamma_R & -i \Delta/2 & i \Delta/2\\
  0 & i \Delta/2 &  -i\Delta/2 & - \Gamma_R/2 & 0\\
    0 & -i \Delta/2 &  i\Delta/2 & 0 & - \Gamma_R/2 
\end{pmatrix}
\rho ,
\end{equation}
In the symmetric case $\Gamma=\Gamma_R=\Gamma_L$, the current and the Fano
factor can be obtained along the lines described in Appendix~\ref{app:FCS} as
\begin{equation}
\label{app:Ifull}
I = \frac{\Gamma \Delta^2}{\Gamma^2+3\Delta^2} ,
\end{equation}
\begin{equation}
\label{app:Ffull}
F = \frac{\Gamma^4 +5\Delta^4-2\Gamma^2\Delta^2}{\left(\Gamma^2+3\Delta^2\right)^2} .
\end{equation}

In the limit $\Delta\ll\Gamma$, considered in the main text, we expand
Eqs.~\eqref{app:Ifull} and \eqref{app:Ffull} to second order in $\Delta$
and obtain
\begin{equation}
I = \frac{\Delta^2}{\Gamma} ,
\end{equation}
\begin{equation}
\frac{C_{2}}{I} = 1-8\frac{\Delta^2}{\Gamma^2}
 = F .
\end{equation}
Moreover, we perform the iteration scheme for the next cumulants within the
same accuracy which provides the expressions
\begin{equation}
\frac{C_{3}}{I} =  1-\frac{24\Delta^2}{\Gamma^2} ,
\end{equation}
\begin{equation}
\frac{C_{4}}{I} =  1-\frac{56\Delta^2}{\Gamma^2} .
\end{equation}
Thus, to lowest order in $\Delta$, all cumulants equal the current,
which indicates that the transport process is essentially Poissonian. 

\bibliography{literature}

\end{document}